# Design of a V-Ti-Ni alloy with superelastic nano-precipitates

J.-L. Zhang[a,b,&], J.L. Cann[a,&], S.B. Maisel[b], K. Qu[c], E. Plancher[a,d], H. Springer[b,e], E. Povoden-Karadeniz[f], P. Gao[c], Y. Ren[g], B. Grabowski[h], C.C. Tasan[a,*]

[a]Dept. of Materials Science and Engineering, Massachusetts Institute of Technology, 77 Mass. Avenue, Cambridge, MA 02139 USA

[b]Max-Planck-Institut für Eisenforschung GmbH, Max-Planck-Straße 1, 40237 Düsseldorf, Germany

[c]International Center for Quantum Materials and Electron Microscopy Laboratory, School of Physics, Peking University, Beijing, 100871, China

[d]Université Grenoble Alpes, CNRS UMR5266, Grenoble INP, Laboratoire SIMaP, 38000 Grenoble, France

[e]Institute of Metal Forming, RWTH Aachen University, 52056 Aachen, Germany

[f]Christian Doppler Laboratory for Interfaces and Precipitation Engineering CDL-IPE TU Wien, Vienna, Austria

[g]X-Ray Science Division, Argonne National Laboratory, Lemont, USA

[h]Institute of Materials Science, University of Stuttgart, Pfaffenwaldring 55, 70569 Stuttgart, Germany

[&]Equal contribution, [*]Corresponding author

**Abstract**

Stress-induced martensitic transformations enable metastable alloys to exhibit enhanced strain hardening capacity, leading to improved formability and toughness. As is well-known from transformation-induced plasticity (TRIP) steels, however, the resulting martensite can limit ductility and fatigue life due to its intrinsic brittleness. In this work, we explore an alloy design strategy that utilizes stress-induced martensitic transformations but does not retain the martensite phase. This strategy is based on the introduction of superelastic nano-precipitates, which exhibit reverse transformation after initial stress-induced forward transformation. To this end, utilizing ab-initio simulations and thermodynamic calculations we designed and produced a $V_{45}Ti_{30}Ni_{25}$ (at%) alloy. In this alloy, TiNi is present as nano-precipitates uniformly distributed within a ductile V-rich base-centered cubic (bcc) β matrix, as well as being present as a larger matrix phase. We characterized the microstructure of the produced alloy using various scanning electron microscopy (SEM) and transmission electron microscopy (TEM) methods. The bulk mechanical properties of the alloy are demonstrated through tensile tests, and the reversible transformation in each of the TiNi morphologies were confirmed by in-situ TEM micro-pillar compression experiments, in-situ high-energy diffraction synchrotron cyclic tensile tests, indentation experiments, and differential scanning calorimetry experiments. The observed transformation pathways and variables impacting phase stability are critically discussed

**Keywords:** martensitic phase transformation; shape memory; NiTi; nanoparticles; in situ

**Graphical Abstract**

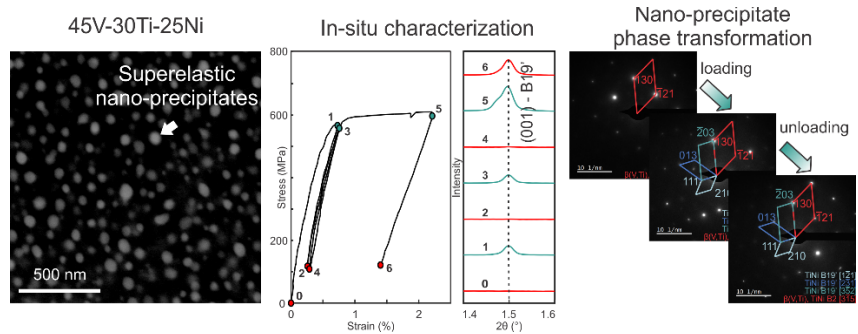

## 1. Introduction

In transformation-induced plasticity (TRIP)-assisted steels, the metastable face centered cubic (fcc) austenite (γ) phase can undergo stress-induced martensitic transformation (γ→α'), enhancing the strain hardening capacity [1,2]. Mechanically-induced martensitic transformation can also increase toughness and fatigue resistance by increasing the energy barrier for micro-crack advancement [3–6]. Thus, this microstructural mechanism has been utilized to improve failure resistance of steels both during forming and in use [7,8]. Moreover, multiple factors that control γ phase stability are being investigated for further optimization of the property benefits (e.g. grain size [9], texture [10], composition [11,12], orientation [2], and dislocation density [13]). TRIP, transformation-toughening, transformation-induced crack closure and other benefits of stress-induced martensitic transformations are not only important for steels. Titanium alloys [14], cobalt alloys [15,16], high entropy alloys [17,18], and various ceramics [19,20] have been reported to benefit from these mechanisms through fcc→body-centered tetragonal (bct), fcc→hexagonal close-packed (hcp), or monoclinic→tetragonal transformations.

A fundamental limitation of the utilization of stress-induced martensitic transformations for improving plastic deformation and damage-resistance is that the newly formed fresh martensite can be damage-prone due to its highly defected substructure [3,21,22]. For example, conventional TRIP-assisted steels with bct martensite show excellent strain hardening, which is beneficial for formability, but their resistance to major causes of failure that occur at low loads or strains like fatigue cracking [4] and hydrogen embrittlement [23] are relatively poor, and post-forming delayed cracking can occur [24]. Similarly, fcc→hcp martensitic transformation also imparts energy-absorptive benefits [25], but brittle hcp martensite remains after the work hardening capability is exhausted, causing quasi-cleavage fracture [26].

Here we investigate a microstructure design strategy based on the introduction of superelastic nano-precipitates within a stable matrix to circumvent this limitation of stress-induced martensitic transformations. By including a superelastic phase, upon removal of the applied stress, the original austenitic microstructure, and thus the original transformation toughening capability can be restored. Thus, rather than aiming to enlarge the shape memory hysteresis or to increase the number of cycles with stable transformation common in shape memory and superelasticity literature [27–30], here we aim to use superelastic constituents to allow for renewable transformation toughening via stress-induced martensitic transformations without producing permanent damage-prone martensite. Previously, the application of superelastic Ti alloys, including TiNi [31,32], Ti-24Nb-4Zr-8Sn [33,34], and Ti-Nb-Zr alloys [35] to fatigue has been investigated, as recoverable elastic strain can accommodate cyclic strain. Prior work on superelastic cold-rolled Ti-24Nb-4Zr-7.6Sn has produced a metastable phase in a combination of nano-size grains and larger grains, increasing its fatigue endurance by ~50% in comparison with the as-forged condition without nano-size grains [33], although that microstructure does not include a stable matrix phase like the one incorporated in this alloy design strategy. One consequence to our approach is that while the most precisely engineered shape memory alloys (such as TiNi, CuAlNi and FeNiCoTi [27]) have particular stoichiometries, we can take more freedom in adjusting alloying elements and dopants to create the multi-phase microstructure that is aimed for. To this end, we design a $V_{45}Ti_{30}Ni_{25}$ (at %) alloy, which, as demonstrated later, has TiNi in two morphologies, as both dispersed TiNi nano-precipitates ($TiNi_{ppt}$) within a stable matrix phase and as a larger TiNi matrix phase ($TiNi_m$). In what follows, we seek to demonstrate whether superelasticity can be achieved in both morphologies present in this alloy, to create the discussed mechanical effects.

Earlier investigations suggest that the behavior of these two morphologies should differ due to the interfacial characteristics, effect of size on stability, and changes in stress and strain partitioning among the phases [36]. For example, in shape-memory nano-precipitates embedded in a coherent metal matrix, the composition of the matrix phase and the lattice mismatch imposed at the phase interface have been shown to influence transformation temperatures [37]. The effect of size on phase stability has been investigated, albeit under different boundary conditions than in the current study. The stability of parent austenite in single-crystalline Cu-14Al-4Ni (wt %) nano-pillars has been found to increase with decreasing pillar diameter by Gómez Cortés et al., who hypothesized that as the pillar size is reduced, the heterogeneous nucleation sites for martensite are also reduced [38]. Molecular dynamics simulations similarly revealed that the martensite start temperature and austenite finish temperature of spherical nanoparticles decreases with decreasing particle diameter [39,40]. The observed correlation between size reduction and austenite stabilization has been attributed to interfacial energy [37,41], shape-strain energy [42], decreased self-accommodation capability [40,43], and decreased number of heterogeneous nucleation sites [44,45].

## 2. Alloy design

To explore this alloy concept, TiNi was selected as the superelastic phase. This choice may not be ideal from a fatigue perspective, due to the softness of martensitic B19' (as compared to austenitic B2) [27] and the high fatigue crack propagation rates reported for superelastic TiNi [31,32]. However, lightly alloyed TiNi exhibits stable superelasticity with a large hysteresis at room temperature [31], making TiNi an ideal phase to achieve a reversible martensitic transformation that can be attained by simply removing the applied stress. In addition, in some ternary systems, TiNi can be introduced coherently within a body-centered cubic (bcc) matrix phase. The introduction of a softer TiNi phase has advantages in limiting ductile damage nucleation. In addition, the martensitic transformation of TiNi enables a transformation toughening effect, which can slow the propagation of cracks. We selected the V-Ti-Ni system, in which TiNi coexists with β, which is a V- and Ti-rich bcc solid solution with a small lattice constant mismatch with TiNi [37], allowing study of the effects of particle size and phase boundary interface on martensitic transformation within a two-phase system with minimal lattice mismatch strain. V-Ti-Ni alloys are already being studied for their potential application as membranes for hydrogen permeation or for hydrogen storage. There have been a number of studies on the effects of changing composition [46–48] and processing parameters [49,50] on alloy microstructure. These alloys are, however, typically brittle and unreliable for application due to intermetallic $Ti_2Ni$ formation [51], which can be difficult to predict as only limited isotherms of V-Ti-Ni ternary phase diagrams are available [52,53].

In this ternary system, it is important to develop a microstructure containing only TiNi and β, but not Ti$_2$Ni. An alloy composition (cyan circle, Fig. 1a$_1$) was chosen within the TiNi + β two-phase region of a 900°C isothermal section of an experimentally-determined ternary phase diagram [52]. After homogenization at 900°C, the microstructure contains a significant volume fraction of brittle Ti$_2$Ni (Fig. 1c). Clearly, consideration of only a single isothermal section is inadequate for the determination of room temperature alloy constitution and composition.

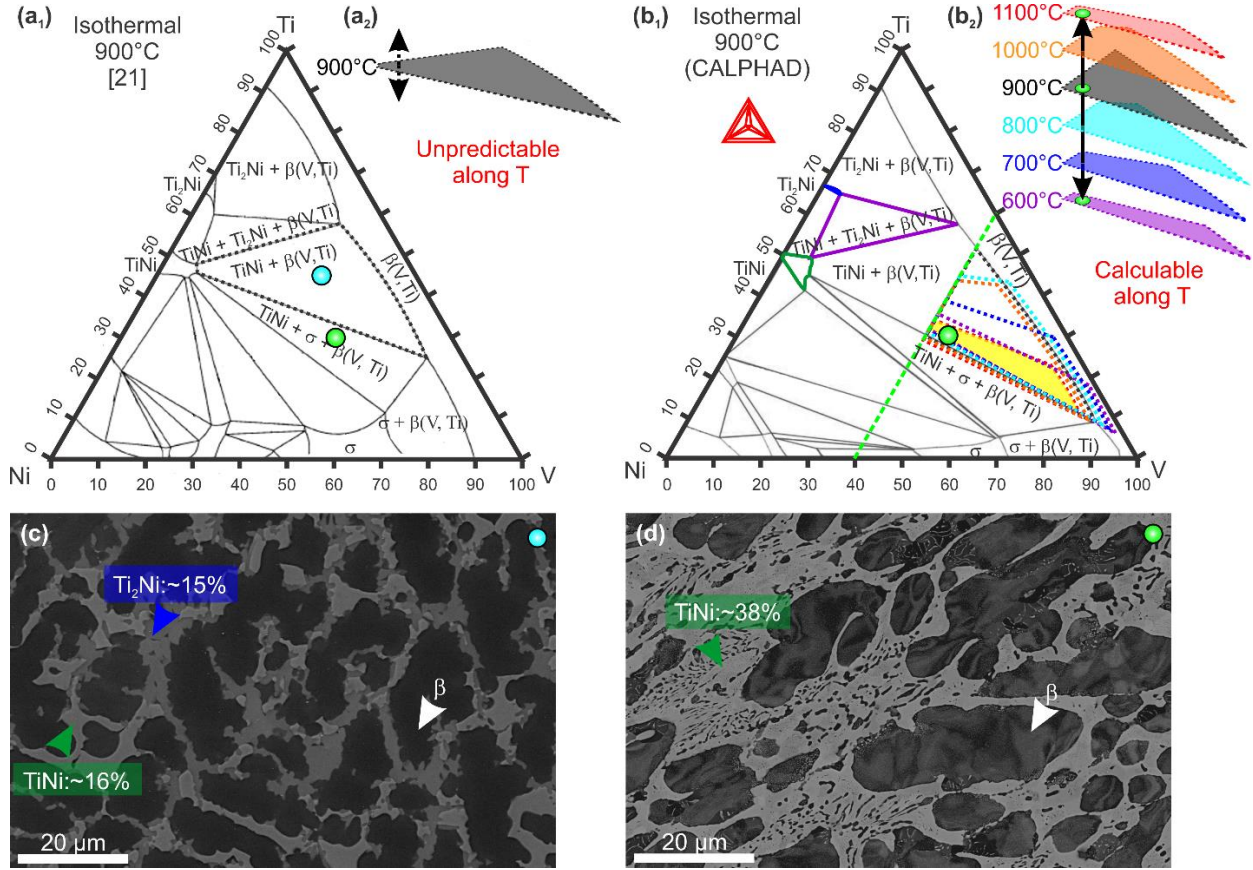

**Fig. 1.** Refinement of the V-Ti-Ni phase diagram. (a$_1$) Isothermal 900°C section of a V-Ti-Ni phase diagram [52]. (a$_2$) Isothermal cross-sections of β-TiNi region of phase diagram [52]. (b$_1$) CALPHAD-based isothermal 900°C section of a V-Ti-Ni phase diagram. (b$_2$) Series of CALPHAD-based isothermal cross-sections of β-TiNi region of phase diagram. (c) Back-scattered electron (BSE) micrograph of microstructure corresponding to the light blue circle in (a). (d) BSE micrograph of microstructure corresponding to the green circle in (a$_1$).

In order to obtain an accurate description of the TiNi + β two-phase region, we have performed high-throughput *ab initio* calculations (Fig. 2). All *ab initio* calculations have been performed using the generalized-gradient approximation [54] as implemented in the VASP package [55,56]. These calculations were done using the projector-augmented wave potentials supplied with VASP [57] at a kinetic energy cutoff of 420 eV. The selection of structures to be calculated has been partially guided by the cluster-expansion approach (Fig. 2, plus signs) [58,59]. Ultimately, a grand total of 186 structures were fully relaxed (Fig. 2, squares) and their resulting energies are shown in Fig. 2. *Ab initio* molecular dynamics simulations within the TU-TILD [60] method were used to validate the results at finite temperatures. We find that bcc-like structures are stable at slightly higher Ni content than suggested by contemporary phase diagrams [52,53]. This suggests that the Ti content must be reduced to remain in the desired two-phase region, avoiding formation of $Ti_2Ni$.

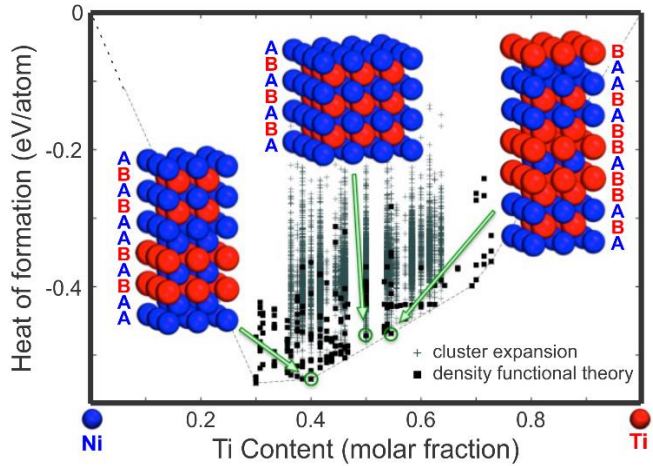

**Fig. 2.** Relaxed structures generated through *ab initio* calculations along with their resulting heats of formation.

Thermodynamic calculations of isothermal sections of the TiNi + β phase field from 600°C to 1100°C in 100°C increments (Figs. $1b_{1-2}$) were carried out using Thermo Calc with a user-specified database, which was created based on both experimental phase stabilities in present literature and first-principle calculations [61]. 1100°C shows the first appearance of the TiNi + β phase field, and 600°C is the low temperature boundary under the assumption that elemental diffusion is insufficient for phase transformation. Dotted outlines of the TiNi + β phase field are shown in Fig. $1b_1$ and in a 3D stack in Fig. $1b_2$. A minimum V content of 40 at. % (Fig. $1b_1$ green dashed line) is considered, as a too large TiNi phase fraction is not desired. Compositions selected from this area should have the desired TiNi + β two-phase microstructure. By overlapping these isothermal sections on the calculated ternary phase diagram, a common area is shown and shaded with in yellow. To avoid an inadequate TiNi phase fraction and problematic metallurgical synthesis due to the high melting temperature of V, $V_{45}Ti_{30}Ni_{25}$ was selected and produced. Its hot rolled and homogenized (2 h at 900°C) microstructure (Fig. 1d) contains the targeted TiNi + β mixture, which is characterized further in Section 4.1.

## 3. Alloy Processing and Characterization

### 3.1. Melt metallurgy and material processing

The test and final alloys with compositions of $V_{35}Ti_{45}Ni_{20}$ and $V_{45}Ti_{30}Ni_{25}$, respectively, were produced via arc-melting under argon atmosphere and cast into a rectangular copper mold. High purity Ni pellets (Ni: >99.9 wt.%), Ti rods (Ti: >99.9 wt.%) and V pieces (V: >99.7 wt.%) were used as raw materials. 200 g ingots were cast, flipped and re-melted in the furnace 6 times to assure chemical homogeneity. The $V_{45}Ti_{30}Ni_{25}$ rectangular ingot was hot rolled at 900°C to ~50% thickness reduction. The hot rolled material was cleaned of its scales by sand blasting, encapsulated in quartz tubes and subsequently annealed for 2 h at 900°C for homogenization, followed by water quenching.

### 3.2. Microstructural characterization

Scanning electron microscopy characterization was conducted using Zeiss Merlin and TESCAN MIRA3 systems. All samples probed by scanning electron microscope (SEM) were wet-ground and polished. Final polishing was carried out using a solution of silica particle suspension with 25% $H_2O_2$. Secondary electron (SE) and backscattered electron (BSE) imaging were conducted using an accelerating voltage of 15 kV. An EDAX/TSL system (AMETEK GmbH) equipped with a Hikari camera was used for electron backscatter diffraction (EBSD) measurements, which were taken with a step size of 50 nm. All EBSD data shown has a minimum confidence index (CI) of 0.1. As both TiNi and β are bcc crystals with only slightly different lattice constants, their Kikuchi patterns are similar, making it difficult to differentiate between them during an EBSD scan. As a result, when the vanadium database was used to index both phases, a distinct variation in image quality parameter (IQ) between the two phases was produced. Bright areas with high IQ are β and dark areas with low IQ are the TiNi phase. Thus, the β and TiNi phases can be partitioned and analyzed individually. This strategy was also confirmed by energy-dispersive X-ray spectroscopy (EDS) mapping to be accurate. EDS was performed at 15 kV with a 10 mm working distance using EDAX TEAM software (AMETEK GmbH). Compositions were determined by averaging the EDS composition from three areas with a size of 25 $\mu m^2$ in each phase.

Transmission electron microscopy (TEM) specimens were lifted out following a site-specific method [62] in a dual-beam focused ion beam (FIB) Helios Nanolab 600i (Thermo Fischer

Scientific). TEM observations were performed in a JEOL JEM-2200 FS (JEOL GmbH) at an operating voltage of 200 kV, through which bright field (BF), dark field (DF) images, and selected area electron diffraction (SAED) patterns were recorded by a Gatan CCD camera (Gatan, Inc.). Scanning transmission electron microscopy (STEM) images were captured using a scanning transmission electron imaging (STEI)-BF detector with a 100 cm camera length. Due to the limitation of the TEM SAED aperture size, additional homogenization treatments to obtain larger nano-precipitates (~130 nm) to facilitate the crystallographic characterization were carried out at 900°C for 48 h on the as-cast and hot rolled materials in a DIL805A/D dilatometer (Bähr Thermoanalyse GmbH), which enables an accurate control of temperature, heating/cooling rate and atmosphere. Argon and helium were used as the protecting atmosphere during holding and as the agent for cooling, respectively.

*3.3. Differential scanning calorimetry*

Differential scanning calorimetry (DSC) tests were carried out using a TA instruments 100 DSC instrument on samples with masses of ~70 mg. A standard Al pan and lid were used to hold the sample. The temperature range was from -100°C to 150°C, and the heating and cooling rates were 10 K·min$^{-1}$. The maximum and minimum temperatures were held for 3 min for stabilization. A test with two empty pans was performed to obtain the baseline, which is applied to each sample data during analysis.

*3.4. Ex-situ mechanical testing*

Dog-bone-shaped tensile samples (gauge geometry: 4×2×1 mm$^3$) were cut by electrical discharge machining (EDM) with the gauge length parallel to the RD. Tensile tests were carried out at strain rate of 10$^{-3}$ s$^{-1}$ using a 5 kN Kammrath & Weiss tensile stage coupled with in-situ imaging using a high speed camera at room temperature. Strain measurements and local strain maps were produced from this data by employing digital image correlation (DIC) analysis using the ARAMIS software (GOM GmbH).

To study the effect of each phase constituent on the crack propagation, interrupted fatigue crack propagation tests were carried out using a Gatan MT2000 tensile stage. The cracks were initiated in an oxide layer at along the edge of the gauge length. This oxide layer was formed when the tensile sample was machined using wire electrical discharge machining (EDM). The room temperature cyclic tests were carried out in a tension-tension mode with minimum and maximum loads of 45 and 440 MPa, respectively (material yield strength: 590 MPa). Crack initiation and propagation at the notch tips were imaged using SE, BSE, and EBSD in a JEOL JSM 6610LV SEM.

Micro-indentation was performed with a load of 100mN with a Vickers tip. Nano-indentation experiments were performed with a Berkovich tip. The load was increased to 250 μN over 30 seconds and was decreased over 30 seconds. Identification of pop-ins was performed according to the methods outlined by Mason et al. [63]. All indentation experiments were performed at room temperature.

*3.5. In-situ mechanical testing*

A synchrotron in-situ tensile experiment was performed at room temperature on beamline 11-ID-C at the Advanced Photon Source at Argonne National Laboratory, operating in transmission mode at 105.7 keV with a 0.5 mm x 0.5 mm beam size and a Perkin Elmer XRD1621 amorphous silicon 2D detector (200 μm pixel size, 2048 x 2048 pixels) with a sample-to-detector distance of 1812 mm. The sample, which was 0.6 mm thick, was secured in a load frame normal to the incident beam. A pattern was drawn onto the sample, enabling DIC to determine the global engineering strain within the gauge length. The sample was strained to $\varepsilon=0.73\%$, subsequently unloaded to $\varepsilon=0.26\%$, loaded again to $\varepsilon=0.76\%$, unloaded to $\varepsilon=0.29\%$, loaded for the final time to $\varepsilon=2.21\%$, and finally unloaded to $\varepsilon=1.40\%$. Each diffraction pattern was obtained from 100 images exposed for 0.65 seconds each. Using the FIT2D software [64,65], the diffraction patterns were integrated from $\varphi=82.5°$ to $97.5°$, as the sample demonstrated a degree of texturing and the diffraction patterns corresponding to the B19' phase were the strongest in this region.

TEM micro-pillar specimens were lifted out following the method by Imrich et al. [66] in a dual-beam Helios G4 FIB (Thermo Fisher Scientific). First, the pillar was shaped to ensure the sides were parallel. Then the sample was milled perpendicular to the compression direction with an overtilt of 1~2° in order to maintain a constant cross-section. The electron-transparent dimension of the micro-pillar was nominally 200 nm, while the nominal width of the micro-pillar was approximately 680 nm and the nominal length was 1250 nm. No in-situ annealing was applied. After the micro-pillars were manufactured, TEM characterization was performed in a Tecnai F20 (Thermo Fisher) at an operating voltage of 200 kV with a 480 mm camera length. High angle annular dark field scanning transmission electron microscopy (HAADF-STEM) images were recorded with a Tecnai F20 with a convergence semi-angle of 30 mrad. TEM EDS measurements were made using Oxford X-Max with α angle tilted 15°. In-situ pillar compression experiments were performed in the Tecnai F20 at room temperature as well under the same working conditions. BF images and SAED patterns were recorded with a Gatan OneView Camera. An in-situ Hysitron PI95 pico-indenter with a flat tip was used with a single-tilt holder, so zone axis selection was limited. Micro-pillar alignment was monitored with bright field to ensure the surface of the micro-pillar was parallel with the indenter surface. The experiment was performed in displacement control mode with a displacement rate of 1 nm/s. There were both loading and unloading segments, with a maximum displacement of 50 nm before returning to the initial position. During the experiment, SAED patterns were recorded with a rate of 0.3 s per frame. Lattice parameters calculated from synchrotron diffraction data were used to index SAED patterns using the SingleCrystal 3.1 software.

## 4. Results
### 4.1. Microstructural characterization

The microstructure of the designed alloy from the micro- to nano-scale is shown in Fig. 3. In the IQ-overlaid phase map (Fig. 3a) obtained from an EBSD scan of the hot rolled bulk sample surface, TiNi, β, and $Ti_2Ni$ are depicted in white with low IQ (darker), white with high IQ (lighter), and blue, respectively. The high-angle grain boundaries (HAGB) are labeled in red, from which it can be seen that the as-cast microstructure has been recrystallized during the hot rolling process. Elemental partitioning between the β ($V_{83}Ti_{12}Ni_5$) and $TiNi_m$ phases are quantified through SEM EDS maps of V, Ti and Ni (Figs. $3b_1$, $b_2$ and $b_3$, respectively). The $TiNi_{ppt}$ (Fig. 3c) are mostly spherical and are homogeneously distributed with an average size of ~50 nm (Fig. $3d_1$). TEM EDS maps of the $TiNi_{ppt}$ (Figs. $3d_2$ to $3d_4$) show that the $TiNi_{ppt}$ are enriched in Ti and Ni and depleted in V.

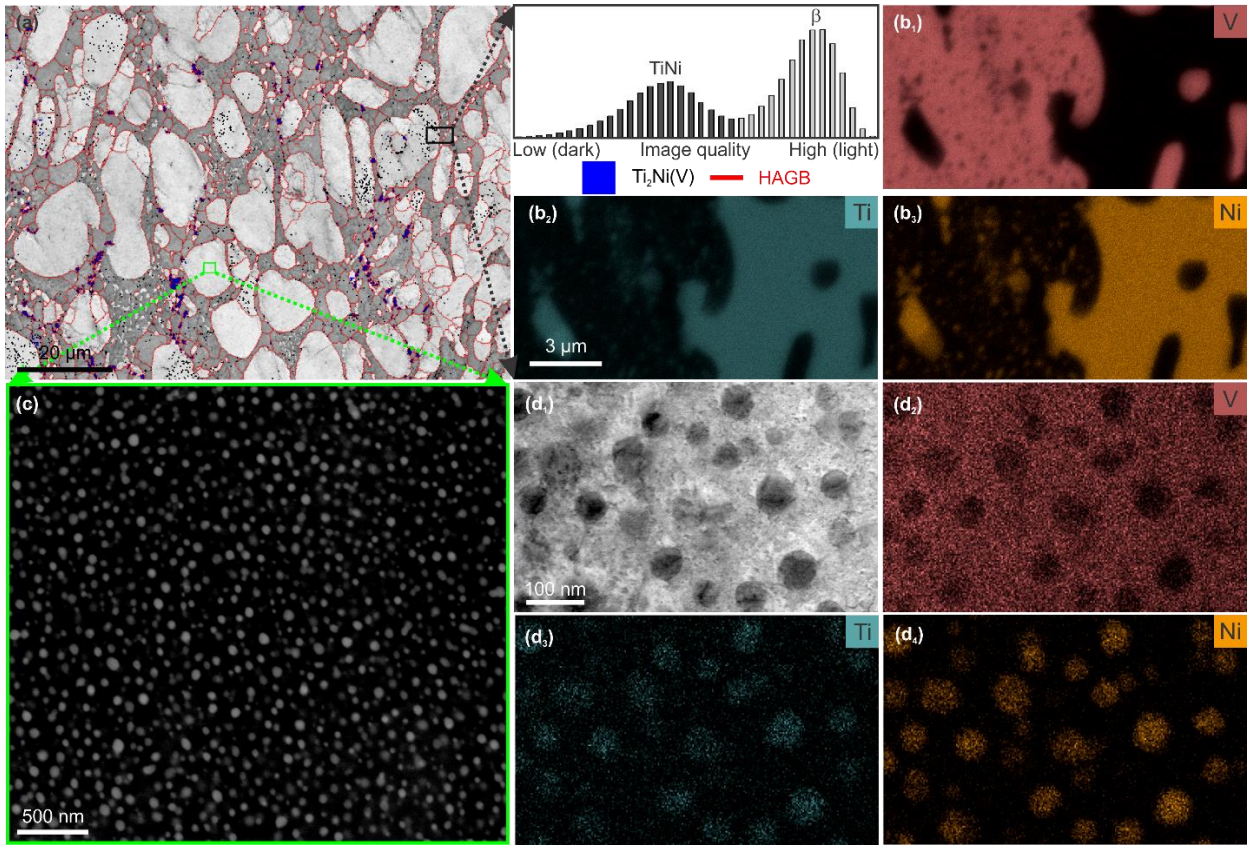

**Fig. 3.** Microstructural characterization. (a) EBSD IQ-overlaid phase map with IQ histogram from entire EBSD map. ($b_1$-$b_3$) EDS analysis. (c) BSE micrograph of TiNi$_{ppt}$ phase. ($d_1$) TEM bright field micrograph. ($d_2$-$d_4$) EDS analysis of TiNi$_{ppt}$.

In order to study the composition and crystallography of the TiNi$_{ppt}$, TEM studies were carried out on a V$_{45}$Ti$_{30}$Ni$_{25}$ sample which has been annealed for 48 hours at 900°C in order to coarsen the TiNi$_{ppt}$ for better observation. The SEM-BSE and TEM BF micrographs are shown in Figs. 4a and 4b$_1$, respectively. TEM EDS (Figs. 4b$_2$-4b$_4$) analysis provides the averaged chemical composition of the TiNi$_{ppt}$ (Ti$_{46.5}$Ni$_{43.7}$V$_{6.2}$O$_{3.6}$). The high O content is likely to be due to the oxidation of the lamella surface (sample not kept in protective environment).

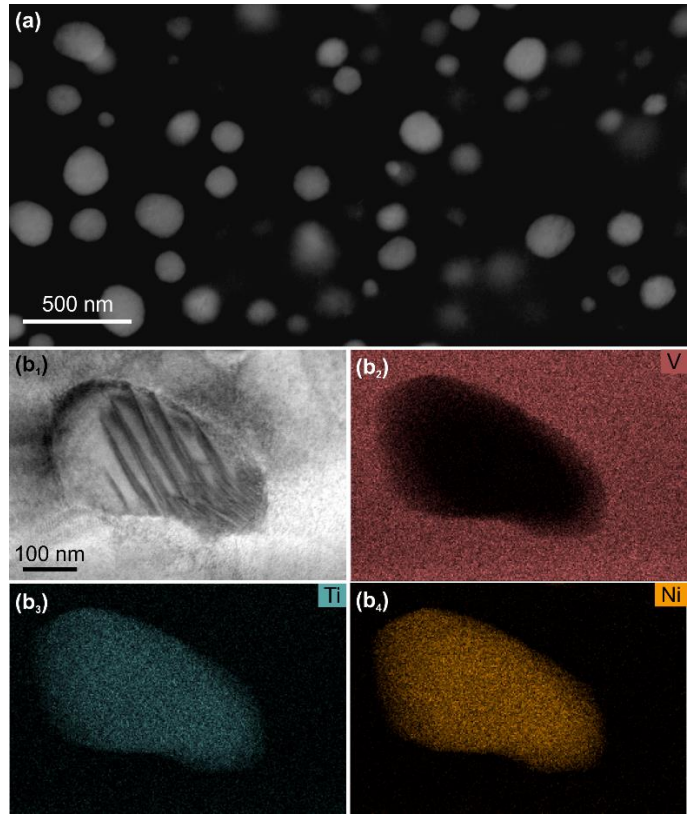

**Fig. 4.** Characterization of grown TiNi$_{ppt}$ in the material treated at 900°C for 48 h. (a) BSE micrograph of grown TiNi$_{ppt}$. (b$_1$) TEM BF micrograph of a selected nano-precipitate. This precipitate appears to contain twins (that may be martensitic), but this was not confirmed with selected area diffraction. The chemical composition averaged from 4 nano-precipitates is Ti$_{46.5}$Ni$_{43.7}$V$_{6.3}$O$_{3.6}$. (b$_2$-b$_4$) TEM EDS analysis of the nano-precipitate shown in (b$_1$).

*4.2. Thermal stability*

Fig. 5 shows the differential scanning calorimetry (DSC) curve of the V$_{45}$Ti$_{30}$Ni$_{25}$ alloy in comparison with a standard Nitinol shape memory alloy (Ti$_{50}$Ni$_{50}$). Both the austenite to martensite transformation and the reverse transformation in the V$_{45}$Ti$_{30}$Ni$_{25}$ alloy separate into two stages (B2 to rhombohedral R-phase and R-phase to monoclinic B19' [67]), whereas this is a one stage transformation in Ti$_{50}$Ni$_{50}$. Most importantly, all transformations shift toward lower temperatures, which are below room temperature, and the austenite finish temperature is only slightly higher than the martensite start temperature.

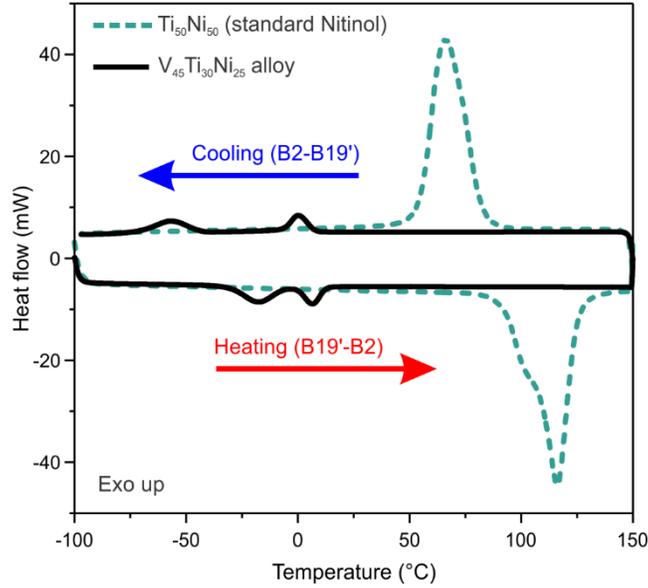

**Fig. 5.** DSC comparison between $Ti_{50}Ni_{50}$ and $V_{45}Ti_{30}Ni_{25}$.

*4.3. Mechanical characterization*

The mechanical properties of $V_{45}Ti_{30}Ni_{25}$ alloy are determined by uniaxial tensile tests. The engineering stress-strain curves are presented in Fig. 6a$_1$. This alloy achieves a balance of high yield stress (590 MPa), ultimate tensile strength (UTS, ~900 MPa), and ductility (~30% tensile elongation). A DIC map of the local strain in the sample gauge is shown in Fig. 6a$_2$. The strain hardening coefficient ($d\sigma/d\varepsilon$) and true stress are plotted against true strain in Fig. 6b. There are multiple distinct stages to the deformation behavior, which according to previous studies [68,69] can be divided as follows: (i) initial elastic loading of the β + B2 TiNi, (ii) stress-induced martensitic transformation with martensite reorientation and detwinning, (iii) elastic loading of martensitic TiNi and plastic deformation of β + remaining B2 TiNi[1], and (iv) plastic deformation of all phases. The critical stress levels for slip β and B2 TiNi in this alloy are unknown but are anticipated to be lower than that of martensitic TiNi. If the TiNi within this alloy is superelastic, martensite transformation in region (ii) should be reversible.

---

1 The transition between regions (ii) and (iii) in Fig. 6a1 occurs at the maximum rate of change in strain hardening coefficient with respect to true strain. This distinction is subtle, however, and thus these two regions are sometimes combined [69].

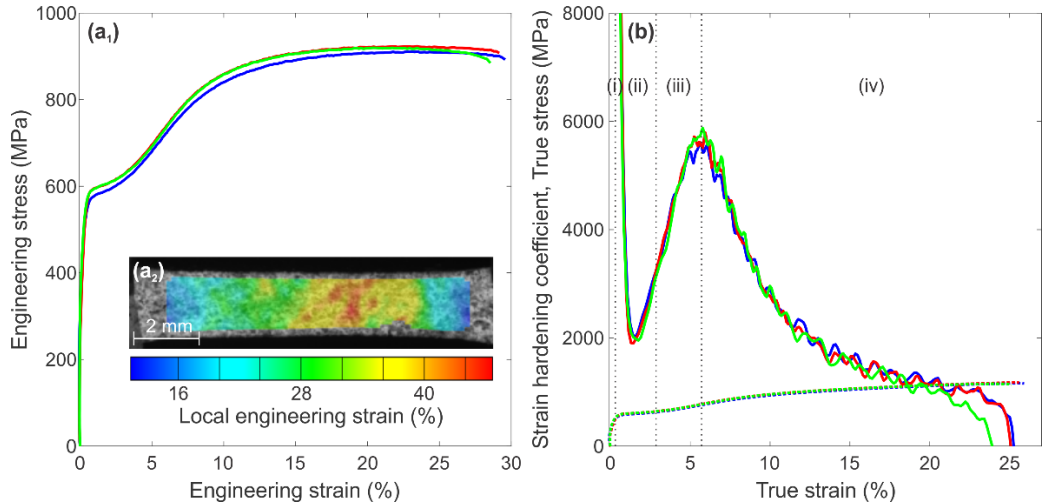

**Fig. 6.** Mechanical behavior of V-TiNi alloy. (a₁) Engineering tress-strain curve from 3 tensile samples, each plotted in a separate color. (a₂) DIC map of local strain levels just prior to fracture. (b) Strain hardening coefficient (solid line) and true stress (dotted line) vs true strain.

An in-situ cyclic synchrotron tensile experiment was performed to test the occurrence of forward and reverse transformations upon straining. The sample underwent three loading-unloading cycles, as shown in Fig. 7a₁. The first two loading-unloading cycles are to an engineering strain (ε) of 0.73%. This strain is larger than that of the first point at which B19' is detected (point A, coinciding with the beginning of region (ii) of Fig. 6b), past which there is a decrease in the strain hardening coefficient due to the martensitic transformation. Figs. 7a₂ and 7a₃ show that upon being strained to ε=0.73% (stage 1), a diffraction spectrum corresponding to the B19' phase appears, and the B2 diffraction spectrum shrinks. Upon unloading (stage 2), the B19' phase disappears, and the B2 phase grows again. These results confirm that at low strains the martensitic transformation in TiNi is reversible. These results were reproduced upon the subsequent loading and unloading cycles (stages 3 and 4, respectively), demonstrating the repeatability of the martensitic transformation. The last loading-unloading cycle is to ε=2.21%, which is near the end of region (ii) of Fig. 6b. At this higher strain (stage 5), a larger proportion of B2 is transformed to B19', and upon unloading (stage 6) the back transformation is not complete. Additionally, this experiment shows that while the thermally-induced martensitic transformation, described in Fig. 5, takes place in two steps (B2 → R → B19'), the stress-induced martensitic transformation takes place in one (B2 → B19'). This will be discussed later.

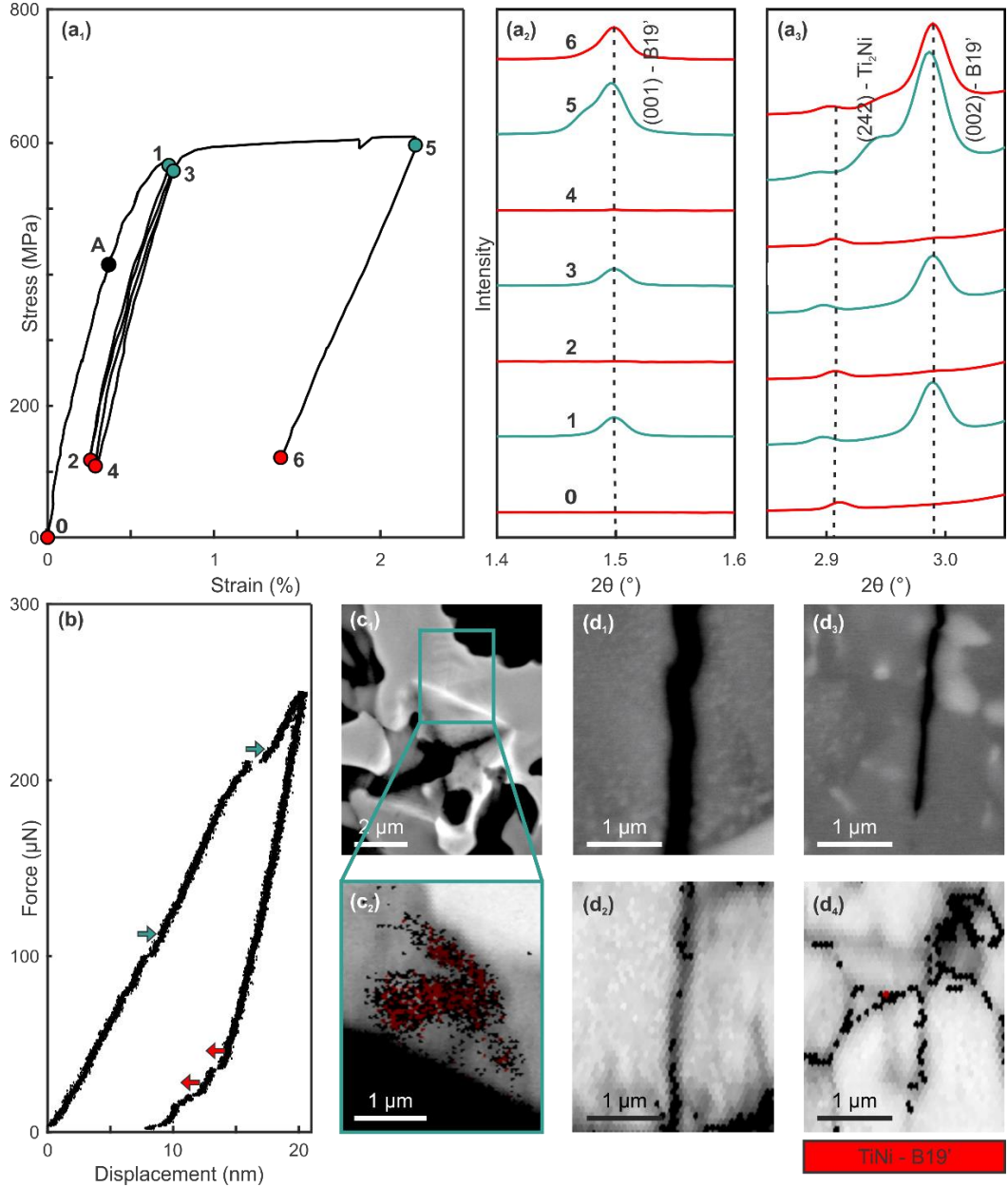

**Fig. 7.** Mechanically-induced forward and reverse transformation. (a) Diffraction spectra at various loading (teal) and unloading (red) states, showing reversible martensitic transformation. B19' peaks first appear in the diffraction pattern at ~410 MPa (point A, black). (b) Load-displacement indentation curve showing pop-ins (teal) and pop-outs (red) from a nanoindent in a β + TiNi$_{ppt}$ region. (c$_{1-2}$) BSE micrograph and EBSD IQ-overlaid B19' phase map of an indent in TiNi$_m$ created through microindentation. (d$_{1-2}$) SE micrograph and EBSD IQ-overlaid B19' phase map of the wake of a fatigue crack. (d$_{3-4}$) SE micrograph and EBSD IQ-overlaid B19' phase map of the tip of a fatigue crack.

In order to show that transformation occurs in TiNi$_m$, a micro-indentation experiment is performed. Transformation of the TiNi$_m$ was captured through backscattered electron (BSE) imaging and EBSD, presented in Fig. 7c$_1$ and Fig. 7c$_2$, respectively. In the BSE image (Fig. 7c$_1$), this can be seen from the contrast change in the BSE image within the TiNi$_m$ grain at the edge of the imprint of the indenter tip. Fig. 7c$_2$ shows an EBSD phase map with B19' (identified in red) indexed in the same location as the contrast change in the Fig. 7c$_1$. The indexed B19' has an average CI of 0.20, and the orientation of the indexed B19' is consistent from point to point.

Demonstrating transformation in TiNi$_{ppt}$ is more challenging. We will describe two experiments providing indirect evidence and one direct proof of phase transformation in TiNi$_{ppt}$. The first indirect evidence comes from nano-indentation experiments performed in the β phase containing TiNi$_{ppt}$, such that the size of the indents is comparable to the size of the nano-precipitates. Two discrete pop-in events (sudden displacement jumps) can be seen in the loading portion of the force-displacement curve (teal arrows, Fig. 7b$_2$). There are also two pop-outs in the unloading portion of the curve at lower forces (red arrows, Fig. 7b$_2$). Pop-ins are frequently used to identify phase transformation in a range of materials [70,71]. In superelastic alloys, pop-outs have been identified to indicate reverse phase transformation [72], where the pop-out loads are lower than the pop-in loads, as observed here. Note, however, that pop-ins can also indicate dislocation activity, shear localization, and brittle fracture [73].

The second indirect evidence is observed from micro-crack propagation experiments carried out to probe the reversibility of the phase transformation of TiNi$_{ppt}$ near crack tips. These micro-crack propagation experiments are not intended to substitute standard fatigue tests – demonstrating microstructure effects on fatigue performance require a dedicated investigation. The micro-crack propagation was analyzed using SE micrographs and EBSD maps in the wake and tip of a fatigue crack (Fig. 7d). SE images reveal that the region in Fig. 7d$_1$ has small nano-precipitates, while those in the crack wake in 7d$_3$ are larger. EBSD maps in the same regions (Fig. 7d$_2$ and 7d$_4$, respectively) both show little to no B19' remaining after unloading. The one point within one of the larger nano-precipitates that indexed as B19' in Fig. 7d$_4$ has a relatively low CI (0.11), and is located at a low angle grain boundary. Note that with the same acquisition and indexing parameters, EBSD does capture B19' in other areas of high stress, as occurred in the vicinity of a fractured brittle Ti$_2$Ni grain (SFig. 1).

In order to more directly demonstrate the phase transformation of TiNi$_{ppt}$, an in-situ micro-pillar compression test was performed in the TEM. Fig. 8a shows a BF micrograph of the micro-pillar aligned with the compression tip. Fig. 8b$_1$ shows a HAADF-STEM micrograph of the micro-pillar. It can be seen in Figs. 8b$_2$-8b$_4$ that there is a large TiNi$_m$ region at the base of the micro-pillar and a β region containing TiNi$_{ppt}$ at the end of the micro-pillar. The stress-strain results of the compression test are given in Fig. 8c. It should be noted that the stresses and strains reported are not exactly those experienced by the TiNi$_{ppt}$, due to stress and strain partitioning. SAED patterns were taken from the nano-precipitate containing region circled in Fig. 8a before loading (Fig. 8d$_1$), at max loading (Fig. 8d$_2$), and after unloading (Fig. 8d$_3$). The SAED patterns were indexed using the lattice parameters determined from synchrotron diffraction. The initial undeformed SAED pattern in Fig. 8d$_1$ can be indexed to austenitic B2 TiNi (a = 3.02 Å), which is the expected phase from EDS analysis. However, the nano-precipitate is not large enough that it is expected to be through-thickness, so there is expected to be a contribution from A2 β as well (a = 3.05 Å). These lattice parameters are similar, and thus contributions from B2 TiNi and A2 β cannot be distinguished down the selected zone axis. The deformed SAED pattern in Fig. 8d$_2$ shows diffraction spots from both the initial phase and B19' (P 1 1 2$_1$/m, a = 2.88 Å, b = 4.52 Å, c = 4.22 Å, γ = 93°). The zone axis of B19' is not precisely the [1 $\bar{2}$ 1], [2 $\bar{3}$ 1], or [3 $\bar{5}$ 2] shown in Fig. 8d$_2$, but a nearby zone. There are also three diffraction spots (grey arrows, Fig. 8d$_2$) that B19' and the initial B2 TiNi/A2 β do not account for. Fig. 8d$_3$ shows the return of the original bright diffraction spots in addition to very weak B19' spots, confirming back-transformation in the TiNi nano-precipitates. A video of the complete evolution of the SAED patterns during deformation can be seen in SFig. 2.

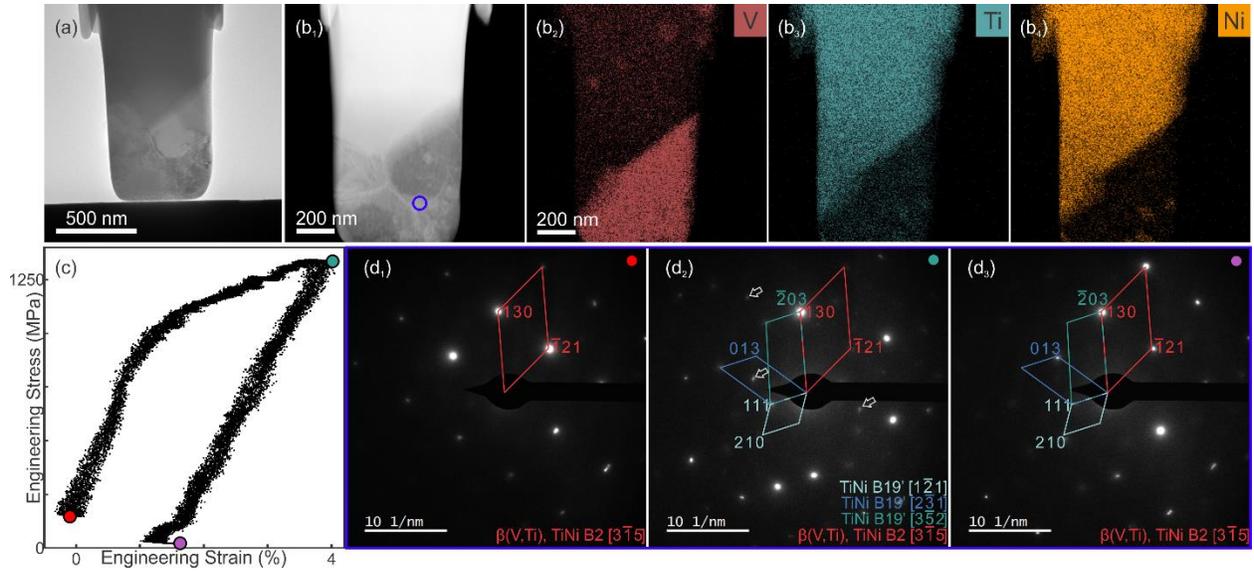

Fig. 8. TEM in-situ micro-pillar compression. (a) BF image of micro-pillar in contact with indenter edge. (b$_1$) HAADF-STEM image of micro-pillar. (b$_{2-4}$) TEM EDS mapping (V, Ni, Ti). (c) Engineering stress-strain plot. (d$_{1-3}$) SAED patterns of the largest nano-precipitate in the micro-pillar, which is circled in blue in (b$_1$). (d$_1$) is the initial state before loading, which corresponds to the red dot in (c). (d$_2$) is at peak load, corresponding to the teal dot in (c). (d$_3$) is after unloading, corresponding to the purple dot in (c).

## 5. Discussion

In the following, we will discuss the deformation response of the TiNi phase that is present in both the TiNi$_{ppt}$ and TiNi$_m$ morphologies, the phase transformation pathways observed, the impact of V-content on alloy properties, the mechanical properties achieved, and others aspects of the alloy.

*5.1 Confirmation of the local superelastic response*

At the bulk-scale, the first demonstration of superelasticity can be derived from the tensile behavior. Previous studies [68,69] have described the superelastic plateau, where the strain-hardening rate approaches zero. V$_{45}$Ti$_{30}$Ni$_{25}$ shows such a region with low strain-hardening rate (Fig. 6b. region ii). The low strain-hardening which does occur in this region may be due to the continued deformation of the β phase and untransformed B2 TiNi. The differential scanning calorimetry experiment (Fig. 5) offers more explicit proof of room temperature superelasticity in this alloy. Finally, direct observation of both forward and reverse transformations were achieved during the in-situ synchrotron cyclic tensile experiments.

These bulk-scale results do not indicate whether it is TiNi$_{ppt}$, TiNi$_m$, or both which exhibit superelasticity. EBSD analyses following micro-indentation experiments demonstrate the ability of TiNi$_m$ to transform (Figs. 7c$_{1-2}$), but more care is needed to confirm transformation in the TiNi$_{ppt}$. A first indication for the ability of the TiNi$_{ppt}$ to undergo martensitic transformation can be sensed from the BF TEM micrograph of a TiNi nano-precipitate (Fig. 4b$_1$), showing the presence of both coarse and fine martensitic twins, along with an untwinned portion. This deformation effect is likely to be due to the slight bending of the TEM lamella during the sample preparation by FIB. A better controlled (yet still indirect) indication is from nanoindentation experiments, which revealed pop-ins in the force-displacement curves on loading, along with pop-outs upon unloading (Fig. 7b). In this case, the presence of pop-outs strengthens the argument that phase transformation is a cause of the pop-ins, as the other mechanisms listed earlier are not reversible. The strongest proof for the transformation tendencies of the TiNi$_{ppt}$, however, is achieved from the in-situ TEM experiments. The TEM micro-pillar compression test in Figs. 8d$_1$-d$_3$ shows the appearance of diffraction spots indexed to B19' with loading and their fading after unloading. Based on these observations, we conclude that the designed alloy exhibits the intended microstructure characteristic, i.e., deformation-induced forward transformation in a second phase, which reverts back upon unloading.

There are, however, a few experimental points worthy of consideration regarding the in-situ TEM compression tests. The first of these is the difficulty in attributing the diffraction pattern in Fig. 8d$_1$ to either B2 or β, as the lattice parameters of these two phases are similar. We propose that the diffraction pattern is due to the superposition of both, as the nano-precipitate diameter is less than the thickness of the micro-pillar. Thus, we cannot conclude what proportion of the nano-precipitate transformed from B2. Secondly, there are also three diffraction spots (grey arrows, Fig. 8d$_2$) which remain unindexed after loading. These spots cannot be indexed to B2 TiNi, β, or B19' in the orientations already observed. They can, however, be indexed to B2 TiNi, β, B19', or even R-phase in another orientation. However, without further spots to confirm which is correct, we cannot decisively conclude which phase these correspond to, though we hypothesize that they may index to the β phase in the neighboring grain. As the pillar compresses throughout the duration of the experiment, there is a corresponding movement of the pillar relative to the aperture opening, with a peak displacement of 50nm from the initial position. As the targeted nano-precipitate is located next to a grain boundary, the signal at maximum displacement may also include signal from the neighboring grain. The third point to consider is that some of the B19' diffraction spots remain after unloading the micro-pillar (Fig. 8d$_3$). This indicates that the reverse transformation is incomplete. While this experiment probes the transformation behavior of only TiNi$_{ppt}$ and reflects the behavior of a smaller, less representative sample, this result is consistent with the results from the bulk synchrotron in-situ cyclic tensile experiments, which also show incomplete reverse transformation.

*5.2 Phase transformation pathway*

Our results show that when martensitic transformation is thermally induced in this alloy, the TiNi phase undergoes a two-stage B2→R→B19' transformation. The same transformation pathway is observed in other V-containing TiNi alloys [74]. However, the synchrotron diffraction experiment shows that when transformation is mechanically induced, V$_{45}$Ti$_{30}$Ni$_{25}$ undergoes a one stage B2→B19' transformation. In order to understand this phenomenon better, we must consider the effect of transformation strain on strain energy in this multi-phase alloy. The transformation strain between B2 and R is approximately 1%, whereas the transformation strain between B2 and B19' is much larger at approximately 10% [27]. Upon cooling, there is a minimal strain between the V and TiNi phases due to coefficient of thermal expansion (CTE) mismatch (CTE$_V$ = 8.4*10$^{-6}$ K$^{-1}$ [75], CTE$_{B2}$ = 13.1*10$^{-6}$ K$^{-1}$ [76]), and there should therefore be minimal additional coherency strain effects on phase stability. Thus, upon cooling, there is not an external strain-induced driving force to promote the formation of B19' over R. There is, however, a larger external strain during tensile loading. Martensitic phase transformation accommodates this external strain, reducing elastic strain energy. This energetic reduction is greater for B19' than R due to the difference in transformation strain. If large enough, the energy saved from reducing strain energy could promote direct B19' formation without an intermediate R-phase. While transformation strains are smaller under compression [77], the same energetic argument can be made for the one-stage B2→B19' transformation observed in the TEM micro-pillar compression experiments. This direct transformation to B19' should be more beneficial from a transformation toughening perspective, as the phase with a lower transformation strain (and thus lower transformation toughening capability [78]) is avoided.

*5.3 Mechanical properties*

The tensile properties of the $V_{45}Ti_{30}Ni_{25}$ alloy show a balance of high yield stress (590 MPa), high UTS (~900 MPa), and high ductility (~30% tensile elongation). The demonstrated strength–ductility combination confirms that this alloy can be formed and sustain significant structural loads. By incorporating a TiNi phase, the alloy is able to exhibit a superelastic plateau, preserving the superelasticity imparted by TiNi and to improve the mechanical response as compared to V-alloys [79] (similar to the matrix phase). It should be noted that the stress-strain curve obtained from the micro-pillar compression experiment does not match the tensile stress-strain curve. This is in large part due to sample size effects, since the micro-pillar sample is not representative of the bulk microstructure, with different phase fraction and texture. But effects of tension-compression asymmetry exhibited by superelastic materials [77] can be in play, as well.

The mechanical properties achieved are due not only to the β and TiNi present, but also to the reduction of the $Ti_2Ni$ fraction in the alloy. By maximizing the temperature range at which the β + TiNi two-phase region is stable (and thus, the range over which $Ti_2Ni$ is not stable), after hot rolling and subsequent annealing at 900°C, the achieved phase fraction of $Ti_2Ni$ is quite small (~1% from EBSD). This could help to reduce brittle failure as a result of $Ti_2Ni$, one of the main challenges to the implementation of V-Ti-Ni alloys as a hydrogen membrane or for hydrogen storage [46,47,80].

A potential application for the transformation toughening effect that resets itself is in increasing the fatigue resistance of structural materials. We observed no martensite at the crack tip or in the crack wake after cyclic tensile deformation (Fig. 7$d_1$-$d_4$). One would expect that due to the high local stresses present at a crack tip, martensitic transformation should occur there. Then, if the martensite is suitably unstable, it might transform back into austenite, re-enabling the transformation toughening mechanism. The lack of martensite detected can be due to one of three options: (1) detection limitations prohibit detection of martensite, even if present, (2) failure to activate the transformation toughening mechanism at all, or (3) successful activation of the transformation toughening mechanism and subsequent back transformation both at the crack tip and in the crack wake. The simplest explanation comes from experimental detection limits. The $TiNi_{ppt}$ in this alloy have an average diameter of ~50 nm, pushing the resolution limit of EBSD, which is approximately 50 nm [81], so transformation may not be detected. The second explanation for the lack of observed martensite is that martensite is not formed at the crack tip during high-cycle fatigue. However, in this alloy, we observe that even at low strains, martensitic transformation can be induced (Figs. 7a$_1$-a$_3$). Nevertheless, the results from the tensile synchrotron experiment are not directly applicable to a crack propagation experiment, as the stress triaxiality is not the same in both scenarios. It has been argued that the positive hydrostatic stress at the crack tip in plane strain conditions could suppress martensitic transformation [82]. However, more definitive in-situ synchrotron diffraction experiments on fatigue-loaded compact tension superelastic TiNi specimens have proven that stress-induced martensite forms at the crack tip in these conditions, and that this transformation is reversible [83]. Thus, it is unlikely that transformation is not induced at all.

*5.4 Alloy design considerations*

The selection of V as the third element in this alloy was partially motivated by the expected coherency between the β phase and the TiNi phase. This was intended to reduce the likelihood of microcrack formation at phase boundaries and to reduce the misfit strain around the nano-precipitates. Previous simulations of pure TiNi nano-precipitates within a TiV β matrix calculate lattice constants of 3.21 Å and 3.02 Å for Ti$_{65}$V$_{35}$ and TiNi, respectively, with lattice mismatch decreasing with increasing V content [37]. Indeed, with lattice parameters obtained through synchrotron experiments of 3.05 Å and 3.02 Å for β and B2 TiNi, respectively, the misfit strain in this alloy should be small and the phases should be well-oriented with respect to one another. Supporting evidence can be found in the micro-pillar compression experiment, in which the diffraction signal came from an area that should contain both B2 TiNi and β before deformation. There are not two distinct patterns, suggesting that if both patterns were present, their orientations were aligned and the superpositioned patterns overlap.

The addition of V, however, affects the lattice parameters not only of the β phase, but also of the martensitic B19' phase. The reported monoclinic angle in pure TiNi B19' ranges from 96.8° to 97.8° [27]. However, with added V content, the monoclinicity decreases [84]. Our results are in line with this finding, with a measured monoclinic angle of 93°. The change in lattice parameters with V addition, however, does not have a significant effect on the unit cell volume. From our experimentally determined lattice parameters, we can calculate the unit cell volume of B19' to be 0.0549 nm$^3$, which is similar to the reported unit cell volume of pure TiNi (0.0546 nm$^3$) [27]. While there is not a large change in cell volume, there is an effect on the dilatational and shear components of the transformation lattice deformation matrix, which can be expressed according to the lattice deformation model as

$$\mathbf{B} = \mathbf{R}\bar{\mathbf{B}}\mathbf{R}^T, \tag{1}$$

where $\mathbf{B}$ is the lattice deformation matrix with respect to the B2 coordinate system, $\bar{\mathbf{B}}$ is the lattice deformation matrix with respect to the B19' coordinate system, and $\mathbf{R}$ is the rotation matrix relating the two coordinate systems. Assuming lattice correspondences of $[1\ 0\ 0]_{B19'} - [1\ 0\ 0]_{B2}$, $[0\ 1\ 0]_{B19'} - [0\ 1\ 1]_{B2}$, and $[0\ 0\ 1]_{B19'} - [0\ 1\ \bar{1}]_{B2}$, $\bar{\mathbf{B}}$ of $V_{45}Ti_{30}Ni_{25}$ can be expressed as

$$\bar{\mathbf{B}} = \begin{bmatrix} 1 & \cot(\gamma) & 0 \\ 0 & 1 & 0 \\ 0 & 0 & 1 \end{bmatrix} \begin{bmatrix} \frac{a}{a_0} & 0 & 0 \\ 0 & \frac{b\sin(\gamma)}{\sqrt{2}a_0} & 0 \\ 0 & 0 & \frac{c}{\sqrt{2}a_0} \end{bmatrix}, \tag{2}$$

where the first matrix is the shear contribution, and the second is the dilatational contribution. Substitution of the B19' lattice parameters, a = 2.88 Å, b = 4.52 Å, c = 4.22 Å, $\gamma = 93°$, and B2 lattice parameter, $a_0 = 3.02$ Å, gives:

$$\bar{\mathbf{B}} = \begin{bmatrix} 1 & -0.0525 & 0 \\ 0 & 1 & 0 \\ 0 & 0 & 1 \end{bmatrix} \begin{bmatrix} 0.954 & 0 & 0 \\ 0 & 1.057 & 0 \\ 0 & 0 & 0.988 \end{bmatrix} = \begin{bmatrix} 0.954 & -0.0555 & 0 \\ 0 & 1.057 & 0 \\ 0 & 0 & 0.988 \end{bmatrix}. \tag{3}$$

The rotation matrix, $\mathbf{R}$, is given as

$$\mathbf{R} = \begin{bmatrix} 1 & 0 & 0 \\ 0 & \frac{1}{\sqrt{2}} & \frac{1}{\sqrt{2}} \\ 0 & -\frac{1}{\sqrt{2}} & \frac{1}{\sqrt{2}} \end{bmatrix}, \tag{4}$$

allowing calculation of the lattice deformation matrix with respect to the B2 coordinate system:

$$\mathbf{B} = \begin{bmatrix} 0.954 & -0.0392 & -0.0392 \\ 0 & 1.023 & 0.0345 \\ 0 & 0.0345 & 1.023 \end{bmatrix}. \tag{5}$$

Comparing this to the lattice deformation matrices of pure TiNi (with B19' lattice parameters, a = 2.889 Å, b = 4.120 Å, c = 4.622 Å, $\beta = 96.8°$, and B2 lattice parameter, $a_0 = 3.015$ Å [27]), given as

$$\bar{\mathbf{B}} = \begin{bmatrix} 1 & 0 & -0.119 \\ 0 & 1 & 0 \\ 0 & 0 & 1 \end{bmatrix} \begin{bmatrix} 0.958 & 0 & 0 \\ 0 & 0.966 & 0 \\ 0 & 0 & 1.076 \end{bmatrix} = \begin{bmatrix} 0.958 & 0 & -0.128 \\ 0 & 0.966 & 0 \\ 0 & 0 & 1.076 \end{bmatrix} \tag{6}$$

and

$$\mathbf{B} = \begin{bmatrix} 0.958 & 0.0905 & -0.0905 \\ 0 & 1.021 & -0.0550 \\ 0 & -0.0550 & 1.021 \end{bmatrix}, \tag{7}$$

it can be seen that the largest dilatational component of the deformation matrix and the shear component are both greater in pure TiNi than in the TiNi present in $V_{45}Ti_{30}Ni_{25}$. This can be anticipated to lessen the transformation strain and the interfacial energy between B2 and newly-formed B19' in $V_{45}Ti_{30}Ni_{25}$. It should be noted that the lattice deformation model is relatively simplistic and neglects the requirement of a undistorted, unrotated habit plane between B2 and B19', which would necessitate the consideration of twinning and an additional rigid body rotation [27,85–87]. Thus, the transformation strain should be less than the lattice deformation model would suggest [85].

These are both factors which could contribute to stabilizing B19', increasing transformation temperatures to some extent. However, experimentally, it has been determined that by adding V to equimolar TiNi, transformation temperatures are decreased [88,89], suggesting that chemical potential is a more important driving force for phase transformation. At 5 at% V, the martensite start temperature ($M_s$) is below room temperature, but the austenite finish temperature ($A_f$) is above room temperature [88]. In $V_{45}Ti_{30}Ni_{25}$ and, more specifically, in the V-containing TiNi phase ($Ti_{46.5}Ni_{43.7}V_{6.2}O_{3.6}$ according to the EDS results in Fig. 4), both $M_s$ and $A_f$ are below room temperature and $A_f$ is only slightly higher than $M_s$ (Fig. 5). These transformation temperatures enable superelasticity at room temperature.

The transformation temperatures may be influenced also by the β matrix-$TiNi_{ppt}$ phase boundary, specifically, the effect of misfit strain and interfacial energy [37]. This effect is size-dependent; the smaller the nano-precipitates are, the lower the transformation temperatures become, past a certain threshold [37]. It should be noted that in Fig. 5, the endo- and exothermic peaks are not bimodal, so the transformation temperatures of the $TiNi_{ppt}$ and the $TiNi_m$ are not distinguishable. This may indicate that the nano-precipitates are not small enough for the size effect to play a significant role. It can be argued that there are two sets of peaks observed, which we attribute to B2 → R → B19' formation, could be attributed instead to the transformation of first the non-dispersed TiNi and then the transformation of the nano-precipitates. However, this is not thought to be likely, as B2 → R → B19' is observed in other V-containing TiNi alloys without the dual TiNi morphologies present in this alloy [74].

The transformation temperatures should have a large impact on transformation toughening. If the transformation toughening mechanism is successfully activated and the microstructure is subsequently totally reset, martensite retention is avoided both at the crack tip and in the crack wake, which can have implications on the extent of fatigue crack toughening. Transformation-induced crack toughing has both intrinsic (ahead of the crack tip) and extrinsic (in the crack wake) components [90]; by reducing the extrinsic component through reverse transformation, the total transformation-induced crack toughening is reduced. This provides another consideration for future alloy design efforts, as transformation temperatures can be used to find a balance between avoiding significant amounts of brittle end product ahead of the crack tip and enhancing the extrinsic component of toughening.

## 6. Conclusions

The goal of this work was to design a multi-phase alloy containing a superelastic phase in order to enable repeated mechanically-induced martensitic transformations upon straining, without retaining a martensitic end phase. This was achieved through the design of the $V_{45}Ti_{30}Ni_{25}$ alloy. This alloy also enables the study of superelasticity in particles confined by a stable metallic matrix and the effects of size on stability in such a system. The main conclusions are presented below:

- The microstructure of the $V_{45}Ti_{30}Ni_{25}$ alloy incorporates TiNi both as nano-precipitates dispersed in a V-rich β matrix and as a larger TiNi matrix phase.
- Through CALPHAD-informed compositional design, the $V_{45}Ti_{30}Ni_{25}$ alloy minimizes brittle $Ti_2Ni$ formation and exhibits at least semicoherency between the nano-precipitates and the β phase. With this composition, a nice combination of strength (yield stress of 590 MPa, UTS of ~900 MPa) and ductility (~30% tensile elongation) is achieved.
- Through a multi-scale experimental campaign, we demonstrated that this alloy exhibits the intended superelasticity in TiNi of both morphologies ($TiNi_{ppt}$ and $TiNi_m$). Bulk-scale experiments confirmed forward and reverse transformation both as a function of stress and temperature. Nano-scale experiments including in-situ TEM compression tests were used to demonstrate the same reversible transformation occurs, even TiNi$_{ppt}$ where size could stabilize the martensitic phase.

- Through synchrotron experiments, the phase transformation was found to be fully reversible at low strain levels, and this phase transformation was repeatable. However, at higher strains, the extent of reverse transformation was decreased.

- While the thermally induced martensitic transformation is found to follow a two-step transformation pathway (B2→R→B19'), the stress-induced martensitic transformation pathway has only one step (B2→B19'). This is hypothesized to be due to the larger transformation strain of B19' in comparison to R.

*Data availability*

The datasets generated and/or analyzed during the current study are available from the corresponding author on reasonable request.

**Acknowledgements**

This work made use of the MRSEC Shared Experimental Facilities at MIT, supported by the National Science Foundation under award number DMR-14-19807. Parts of this work was performed in part at the Center for Nanoscale Systems (CNS), a member of the National Nanotechnology Coordinated Infrastructure Network (NNCI), which is supported by the National Science Foundation under NSF award no. 1541959. CNS is part of Harvard University. This research used resources of the Advanced Photon Source, a U.S. Department of Energy (DOE) Office of Science User Facility operated for the DOE Office of Science by Argonne National Laboratory under Contract No. DE-AC02-06CH11357. The research was partially supported by funding by the Deutsche Forschungsgemeinschaft (SPP 1568) and by the European Research Council (ERC) under the European Union's Horizon 2020 research and innovation program (Grant Agreement No. 639211).  Discussions with S. Wei and B. Hallstedt are acknowledged.

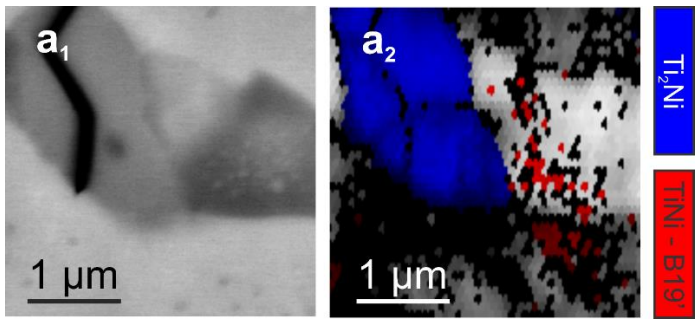

**Supplementary Figure 1.** EBSD observation of B19'. ($a_1$), SE micrograph of region near fatigue crack in brittle $Ti_2Ni$ phase. ($a_2$), EBSD IQ-overlaid phase map showing presence of B19' (average CI of 0.17) in both TiNi nano-precipitates and non-dispersed TiNi.

SFigure 2_Video.mp4

**Supplementary Figure 2.** Video of TEM micro-pillar compression test. It can be noted that there are diffraction spots which appear and disappear before the sample is fully loaded. These diffraction spots also index to B19', and their appearance and subsequent disappearance are hypothesized to be due to crystal rotation or micro-pillar bending during deformation.